\documentclass[letterpaper]{article}
\usepackage{setspace}
\usepackage{color}
\usepackage{amsmath}
\usepackage{amsfonts}
\usepackage{verbatim}
\usepackage{amssymb}
\usepackage{graphicx,bm}
\usepackage{graphicx}
\usepackage{subcaption}
\usepackage{float}

\DeclareMathOperator{\sech}{sech}

\definecolor{darkgreen}{rgb}{0,0.35,0}

\providecommand{\U}[1]{\protect\rule{.1in}{.1in}}
\begin{document}

\title{Analytic non-Abelian gravitating solitons in the
Einstein-Yang-Mills-Higgs theory and transitions between them}
\author{Fabrizio Canfora$^{1}$ and Seung Hun Oh$^{2}$, \\
$^{1}$\textit{Centro de Estudios Cient\'{\i}ficos (CECS), Casilla 1469,
Valdivia, Chile}\\
$^{2}$\textit{Department of Liberal Arts, Korea Polytechnic University, } \\
\textit{\ Sangidaehak-ro 237, Siheung-si, Gyeonggi-do, 15073, Korea}\\
{\small canfora@cecs, shoh.physics@gmail.com}}
\maketitle

\begin{abstract}
Two analytic examples of globally regular non-Abelian gravitating solitons in the Einstein-Yang-Mills-Higgs theory in (3+1)-dimensions are presented. In both cases, the space-time geometries are of the Nariai type and the Yang-Mills field is completely regular and of meron type (namely, proportional to a pure gauge). However, while in the first family (type I) $X_{0} = 1/2$ (as in all the known examples of merons available so far) and the Higgs field is trivial, in the second family (type II) $X_{0}$ is not 1/2 and the Higgs field is non-trivial. We compare the entropies of type I and type II families determining when type II solitons are favored over type I solitons: the VEV of the Higgs field plays a crucial role in determining the phases of the system. The Klein-Gordon equation for test scalar fields coupled to the non-Abelian fields of the gravitating solitons can be written as the sum of a two-dimensional D'Alembert operator plus a Hamiltonian which has been proposed in the literature to describe the four-dimensional Quantum Hall Effect (QHE): the difference between type I and type II solutions manifests itself in a difference between the degeneracies of the corresponding energy levels.
\end{abstract}

\section{Introduction}

Non-Abelian solitons and instantons are fundamental pillars in our
understanding of gauge theories beyond perturbation theories (see \cite{94co}
\cite{3act} \cite{267raj} \cite{mantonbook} \cite{BaMa} and references
therein). In many situations (such as in the cases of topological defects in
the early universe), the effects of gravity cannot be neglected (see \cite%
{vilenkin} and references therein). From a genuine General Relativistic
viewpoint, non-Abelian gravitating solitons represent severe tests for
well-known conjectures such as the no-hair conjecture. Consequently, it is a
mandatory task to shed further light on these types of gravitating solitons.
The gravitating solitons which have been analyzed in more details in the
eighties and nineties are asymptotically flat and (almost always) numerical.
However, it is worth noting that the requirement of asymptotic flatness is
somewhat ``foreign" to the requirement of regularity in the sense that while
gravitating solitons must be regular by definition, genuine non-Abelian and
globally regular gravitating solitons need not to be asymptotically flat%
\footnote{%
Thus, as it is now usual in the literature, in the present manuscript the
notion of ``gravitating soliton" means ``globally regular" but not necessarily
asymptotically flat.}. The first example was discovered by Bartnik and
McKinnon (BK) \cite{19bmk}. Soon after the BK gravitating solitons, genuine
non-Abelian black holes were also constructed numerically in \cite{324vol} 
\cite{215ku} \cite{33bi}. These results closely related to the no-hair
conjecture and to the black holes uniqueness theorem attracted a lot of
attention since then (see, for instance, \cite{GalPhysRep} \cite{win} \cite%
{win0} \cite{win1} \cite{win2} \cite{win3} \cite{win4} and references
therein). In the present paper, we will be mainly interested in the
construction of analytic non-Abelian gravitating solitons. To the best of
our knowledge, the only known analytic example in (3+1) dimensions of a
globally regular gravitating soliton with ``bona fide" non-Abelian gauge
field has been found in \cite{chamvolkov} (many nice numerical examples are
described in \cite{GalPhysRep} \cite{win} \cite{win0} \cite{win1} \cite{win2}
\cite{win3} \cite{win4} and references therein). Such remarkable analytic
solution has been found in the Einstein-Yang-Mills-dilaton system: so far it
has not been possible to extend this solution to Einstein-Yang-Mills theory
without dilatonic coupling. The main goal of the present manuscript is to
construct two types of globally regular analytic non-Abelian gravitating
solitons in the Einstein-Yang-Mills-Higgs system in (3+1)-dimensions without
any dilatonic coupling and to analyze the transitions between these families
as well as their remarkable physical properties.

An obvious question is: why should one insist so much on finding analytic
solutions if these equations can be solved numerically? Since the pioneering
works of Bartnik and McKinnon mentioned above, very powerful numerical
techniques have been proposed in the literature to construct non-Abelian
gravitating solitons (see, for instance, \cite{GalPhysRep}, \cite{48-49} 
\cite{50wa} \cite{51ni} and references therein). There are really good
reasons to strive for analytic solutions nevertheless.

Besides the obvious fact that a systematic tool to construct analytic
gravitating solitons can greatly enlarge our understanding of these
configurations, the main reason is that the availability of analytic
gravitating non-Abelian solitons allows to disclose phenomena that would be
very difficult to see otherwise. In particular, the present formalism
discloses the possibility to have transitions between these solitons and a
very surprising analogy with the four-dimensional Quantum Hall Effect (QHE)
which will be discussed in a moment.

As we are seeking for genuine non-Abelian effects, meron types
configurations (introduced in \cite{alfubini}) are really good candidates as
such configurations can only appear in non-Abelian gauge theories (see \cite%
{3act}\ and references therein). A gauge potential of meron type can be
defined as proportional to a pure gauge: $A_{\mu }=X_{0}U^{-1}\partial _{\mu
}U$ (with $X_{0}\neq 0,1$): such a gauge potential is non-trivial due to the
presence of the commutator in the non-Abelian field-strength. On the other
hand, in Abelian gauge theories, a gauge potential which is proportional to
a pure gauge is itself a pure gauge\footnote{%
In the Abelian case, when $X_{0} $ is constant, $A_{\mu }=X_{0}\partial
_{\mu }\Omega \Rightarrow A_{\mu }=\partial _{\mu }\left( X_{0}\Omega
\right) $.} and therefore is trivial. Consequently, merons are \textit{bona
fide} non-Abelian configurations. Since the pioneering works \cite{meronsCDG}
\cite{meronsjaffe}, the important role of these types of configurations in
understanding the non-perturbative phase of QCD has been widely recognized
(see \cite{meronslenz}\ and references therein).

All the known examples of merons so far have $X_{0}=1/2$; thus, a first
question that we will answer (affirmatively) is: is it possible to have
non-trivial merons configurations with $X_{0}\neq 1/2$ ?

The second and most difficult issue related to merons is the following.
Merons on flat spaces are necessarily singular: they play an important role
(see \cite{meronsCDG} \cite{meronsjaffe} \cite{meronslenz}\ and references
therein) as ``elementary components" of instantons (since an instanton can be
thought as a bound state of two merons). However, on flat spaces, merons
cannot be observed directly (as they have infinite Euclidean action/energy).
When Yang-Mills theory is coupled with General Relativity (GR), it has been
possible to construct analytic examples of merons-black holes \cite{meron0} 
\cite{mimeron1} \cite{mimeron2} \cite{mimeron3}. Thus, in a sense,
meron-black holes can be observed directly (some peculiar effects have been
discussed in \cite{mimeron3}), but the meron singularity is still there
(although hidden behind the horizon). Consequently, the examples in \cite%
{meron0} \cite{mimeron1} \cite{mimeron2} \cite{mimeron3} are not gravitating
solitons but rather non-Abelian black holes. The very important question is: 
\textit{can we construct analytic examples of gravitating merons in which
the typical singularity of the merons disappears completely?} In the
following, we will show that it is indeed possible to construct globally
regular gravitating merons solutions free of any singularity. Moreover,
there are two different families of regular gravitating merons that compete
against each other leading to quite interesting phenomena.

A further by-product of the present analysis appears when one analyzes the
dynamics of a test field charged under the gauge group moving on these
gravitating non-Abelian solitons. The effective Hamiltonian determining the
dynamics of charged test fields includes the Hamiltonian describing the
four-dimensional Quantum Hall Effect (4DQHE) introduced in \cite{4DQHE1} and
further analyzed in \cite{4DQHE2} \cite{4DQHE3} \cite{4DQHE4} \cite{4DQHE5} 
\cite{4DQHE6} \cite{4DQHE7} \cite{4DQHE8}\ and references therein. Such deep
generalization of the usual theory of the two-dimensional QHE has also been
confirmed in condensed matter experiments (see \cite{4DQHEexp1} \cite%
{4DQHEexp2}\ and references therein). However, until now, there have been
very few concrete realizations of the 4DQHE in high energy physics using
fields arising in the standard model of particles physics minimally coupled
to GR. In this work, we fill the gap by providing a setting (in the
Einstein-Yang-Mills-Higgs theory\footnote{%
To the best of the authors knowledge, the first glimpse of the relevance of the
usual QHE in astrophysical Black Holes has been provided in \cite{BHQHE}.})
in which an explicit realization of the physical features of the 4DQHE is
possible. A very intriguing effect is that the two families of non-Abelian
gravitating solitons can be distinguished by looking at the degeneracies of
the corresponding energy levels.

The paper is organized as follows: In Section II, we present the model and
give our ansatz and the corresponding field equations. 
In Sections III and IV, we construct the two families of the analytic regular 
solutions of the Einstein-Yang-Mills-Higgs theory in (3+1)-dimensions. 
In Section V, we introduce a standard coordinate system of the Nariai class 
of spacetime to work out the surface gravity and the temperature of the 
system. In Section VI, we present the entropy functions of the two types of 
the solutions and provide some useful plots of the entropy functions to see
which configuration is favored for given sets of the physical parameters. 
In Section VII, we analyze the Klein-Gordon equation to disclose the 
difference between two types of solutions by making use of the 
four-dimensional quantum hall effects. In the last Section, our conclusions 
are drawn.

\section{Action and ansatz}

The starting point is the action of the (3+1)-dimensional
Einstein-Yang-Mills-Higgs field: 
\begin{equation*}
I=\int d^{4}x\sqrt{-g}\Big(\frac{R-2\Lambda }{\kappa }+\frac{1}{4e^{2}}\text{%
Tr}\big[F_{\mu \nu }F^{\mu \nu }\big]+\frac{1}{2e^{2}}\text{Tr}\big[D_{\mu
}\Phi ^{\ast }D^{\mu }\Phi \big] - 2V(\Phi ^{\ast }\Phi )\Big)\ ,
\end{equation*}%
where $R$, $\Lambda $, and $V(\Phi ^{\ast }\Phi )$ are the Ricci scalar of
the space-time, the cosmological constant, and the self-interacting Higgs
potential, respectively. The dimensionless constant $e$ is the gauge
coupling and $\kappa =8\pi G$ for the gravitational constant $G$. The field
strength of the Yang-Mills field $A_{\mu }$ is $F_{\mu \nu }=\partial _{\mu
}A_{\nu }-\partial _{\nu }A_{\mu }+[A_{\mu },A_{\nu }]$ and the
gauge-covariant derivative is $D_{\mu }=\nabla _{\mu }+[A_{\mu },\cdot \ ]$.
In this notation, the Einstein's equations are written as 
\begin{equation*}
R_{\mu \nu }-\frac{1}{2}g_{\mu \nu }R + \Lambda g_{\mu\nu} =\kappa T_{\mu
\nu }\ ,
\end{equation*}%
where the energy-momentum tensor is given by 
\begin{equation*}
T_{\mu \nu }=T_{\mu \nu }^{\text{(YM)}}+T_{\mu \nu }^{\text{(H)}}\ ,
\end{equation*}%
where 
\begin{align}
T_{\mu \nu }^{\text{(YM)}}& =-\frac{1}{2e^{2}}\text{Tr}\Big(g^{\alpha \beta
}F_{\mu \alpha }F_{\nu \beta }-\frac{1}{4}g_{\mu \nu }F^{\alpha \beta
}F_{\alpha \beta }\Big)\ , \\
T_{\mu \nu }^{\text{(H)}}& =-\frac{1}{2e^{2}}\text{Tr}\Big(D_{\mu }\Phi
D_{\nu }\Phi -\frac{1}{2}g_{\mu \nu }D^{\alpha }\Phi D_{\alpha }\Phi \Big)%
-g_{\mu \nu }V(\Phi ^{\ast }\Phi )\ .
\end{align}%
The equations for the Higgs and Yang-Mills fields are 
\begin{align}
g^{\mu \nu }D_{\mu }D_{\nu }\Phi & = -e^{2} \frac{dV}{d(\Phi ^{\ast }\Phi )}%
\Phi \ ,  \label{Higgs} \\
D^{\mu }F_{\mu \nu }& =[\Phi ,D_{\nu }\Phi ]\ .  \label{ym}
\end{align}%
The Higgs potential is given by 
\begin{equation*}
V(\Phi ^{\ast }\Phi )=\lambda (\Phi ^{\ast }\Phi -v_{0}^{2})^{2}\ ,
\end{equation*}%
where $\lambda $ is the self-interacting coupling constant of the Higgs
fields. We consider this Einstein-Yang-Mills-Higgs (briefly, EYMH) system in
the space-time with the metric given by 
\begin{equation*}
ds^{2}=F_{0}\big[-2y(u,v)dudv+L^{2}\big(d\theta ^{2}+\sin ^{2}\theta \ d\phi
^{2}\big)\big]\ ,
\end{equation*}%
where $L$ and $F_{0}$ are constant (without loss of generality $F_{0}$ can
be assumed to be positive). As has been already emphasized, the
Yang-Mills field is assumed to have the meron form 
\begin{equation}
A_{\mu }=X_{0}\ U^{-1}\partial _{\mu }U\ ,
\end{equation}%
where $X_{0}$ is a constant such that $X_{0}\neq 0,1$. The $SU(2)$-valued
scalar field $U$ is parametrized as 
\begin{equation*}
U^{\pm 1}(x^{\mu })=\pm Y^{A}(x^{\mu })t_{A}\ ;\qquad Y^{1}=\sin \theta \cos
\phi \ ,\,Y^{2}=\sin \theta \sin \phi \ ,\,Y^{3}=\cos \theta \ ,
\end{equation*}%
where $t_{A}=i\sigma _{A}$ for the Pauli matrix $\sigma _{A}$ The Higgs
field is given in an adjoint representation by 
\begin{equation*}
\Phi =W_{0}\ U\ ,
\end{equation*}%
where the constant $W_{0}$ has to be determined solving the Higgs field
equations which (with the ansatz defined above) reduce to the single
algebraic equation 
\begin{equation}
2 \lambda e^{2} F_{0} L^{2} (W_{0}^{2}-v_{0}^{2})+(2X_{0}-1)^{2}=0\ .
\label{algebr1}
\end{equation}%
The Yang-Mills equations \eqref{ym} also reduce to the following single
algebraic equation; 
\begin{equation}
(2X_{0}-1)\big[L^{2}F_{0}W_{0}^{2}-X_{0}(1-X_{0})\big]=0\ .  \label{algebr2}
\end{equation}%
From Eqs. (\ref{algebr1}) and (\ref{algebr2}) it is clear that there are two
families of solutions.

The first family (which will be called \textbf{type I}) corresponds to the
usual meron solution together with the condition that the Higgs field
profile $W_{0}$ is in the ``VEV":%
\begin{equation}
X_{0}=\frac{1}{2}\ ,\ W_{0}^{2}=v_{0}^{2}\ .  \label{typeI}
\end{equation}

The second family (which will be called \textbf{type II}) corresponds to the
conditions%
\begin{equation}
X_{0}\neq \frac{1}{2}\ ,\ W_{0}^{2}\neq v_{0}^{2}\ .  \label{typeII}
\end{equation}%
The explicit form of the type \textbf{II} solution will be discussed in the
next sections.

The nonvanishing components of the Einstein's equation are found to be 
\begin{align}
e^{2} \big[ \kappa \lambda (W_{0}^{2}-v_{0}^{2})^{2} + \Lambda \big] %
(F_{0}L^{2})^{2} + \big[ \kappa W_{0}^{2} (2 X_{0} -1)^{2} - e^{2} \big] (
F_{0} L^{2} ) + 2 \kappa X_{0}^{2}(1-X_{0})^{2} =0\ ,  \label{reduced1} \\
\frac{1}{y}\partial _{u}\partial _{v}y-\frac{1}{y^{2}}\partial _{u}y\cdot
\partial _{v}y + \Big( \frac{2\kappa X_{0}^{2}(1-X_{0})^{2} } {%
e^{2}(F_{0}L^{2})^{2} } - \kappa \lambda (W_{0}^{2}-v_{0}^{2})^{2} -\Lambda %
\Big) F_{0} y =0\ .  \label{reduced2}
\end{align}

\section{Type I family: standard Meronic configuration}

It follows from Eq. \eqref{algebr1} that the value of $W_{0}$ is precisely
the vacuum expectation value $v_{0}$ if and only if the configuration of the
gauge field is a meron: 
\begin{equation*}
W_{0}=v_{0}\qquad \Longleftrightarrow \qquad X_{0}=\frac{1}{2}\ .
\end{equation*}%
In this case, the Eqs. (\ref{algebr1}) and (\ref{algebr2}) are automatically
satisfied and the Eq. (\ref{reduced1}) fixes the value of $F_{0} L^{2}$,
which is the size of the $S^{2}$, in terms of the cosmological constant $%
\Lambda $ and of the other parameters: 
\begin{equation*}
\Lambda (F_{0}L^{2})^{2} - F_{0}L^{2} + \frac{\kappa}{8 e^{2}}=0 \ .
\end{equation*}%
This equation admits one or two positive roots for $F_{0} L^{2}$ and the
number of roots depends on the range of physical parameters as follows: 
\begin{align}
\begin{cases}
F_{0} L^{2} = \frac{1 \pm \sqrt{1-\kappa \Lambda/2e^{2}}}{2 \Lambda} >0
\qquad \text{when} \quad 0 < \Lambda < 2e^{2}/\kappa \ , \\ 
F_{0} L^{2} = \frac{1 - \sqrt{1-\kappa \Lambda/2e^{2}}}{2 \Lambda} >0 \qquad 
\text{when} \quad \Lambda < 0 \ , \\ 
F_{0} L^{2} = \frac{1}{2 \Lambda} \qquad \text{when} \quad \Lambda =
2e^{2}/\kappa \ , \\ 
F_{0} L^{2} = \frac{\kappa}{8 e^{2}} \qquad \text{when} \quad \Lambda =0 \ .%
\end{cases}%
\end{align}

\section{Analytic Solutions of type II}

The Higgs equation \eqref{algebr1} has the solution 
\begin{equation}
X_{0}=\frac{1}{2}\Big(1\pm \sqrt{2\lambda
e^{2}F_{0}L^{2}(v_{0}^{2}-W_{0}^{2})}\Big)\ ,  \label{TypeIIX0}
\end{equation}%
only when the constant $W_{0}$ lies in the range of 
\begin{equation}
W_{0}^{2}\leq v_{0}^{2}\ .  \label{W0const}
\end{equation}%
When $W_{0}=v_{0}$ the configuration of the Yang-Mills field becomes that of
a standard meron $X_{0}=1/2$ (and the Higgs field becomes trivial as it
completely disappears from the energy-momentum tensor) so the solutions
become of type \textbf{I}. In this section, we will discuss $W_{0}$ such
that $W_{0}^{2}<v_{0}^{2}$ (the type \textbf{I} solutions will be discussed
in the next section). The Yang-Mills equation \eqref{algebr2} gives 
\begin{equation}
F_{0}L^{2}=\frac{1}{2\big(\lambda e^{2}(v_{0}^{2}-W_{0}^{2})+2W_{0}^{2}\big)}%
\ .  \label{TypeIIF0}
\end{equation}%
Using \eqref{TypeIIX0} and \eqref{TypeIIF0} in Eq. (\ref{reduced1}), we
obtain a quadratic equation for $W_{0}^{2}$ 
\begin{equation}
\kappa (2-\lambda e^{2})(W_{0}^{2})^{2}-2e^{2}(2-\lambda
e^{2})W_{0}^{2}+e^{2}\big[\lambda v_{0}^{2}(\kappa v_{0}^{2}-2e^{2})+\Lambda %
\big]=0\ .  \label{quadratic}
\end{equation}%
This equation admits one or two positive roots for $W_{0}^{2}$ within $\big(%
0,v_{0}^{2}\big)$. Let us examine the corresponding configurations of the
physical system in order.

\subsection{Type \textbf{II} configurations}

There are three cases that must be considered separately.

\textbf{Option 1)} Eq. (\ref{quadratic}) can have two positive roots for $%
W_{0}^{2}$ (in this case type II configurations can be divided into two
sub-families, one for each positive root of Eq. (\ref{quadratic})).

\textbf{Option 2)} Eq. (\ref{quadratic}) can have one positive roots for $%
W_{0}^{2}$ (in this case, there is just one family of type II configurations).

\textbf{Option 3)} Eq. (\ref{quadratic}) has no positive roots for $%
W_{0}^{2} $ (in this case, there is no family of type II configurations).

Whether (for instance) \textbf{option 1} is realized instead of \textbf{%
options 2} or \textbf{3} depends on the values of the parameters of the
models. Especially relevant are the Higgs coupling constant $\lambda $ and
the VEV $v_{0}^{2}$. \textbf{Option 1} is the most interesting one from the
thermodynamical viewpoint since, in this case, there is a competition
between three types of solutions: the standard meron $X_{0}=1/2$ and $%
W_{0}^{2}=v_{0}^{2}$ (type \textbf{I}), the type \textbf{II} solution
corresponding to the larger positive root of Eq. (\ref{quadratic}) and the
type \textbf{II} solution corresponding to the smaller positive root of Eq. (%
\ref{quadratic}). As it will be discussed in the next sections, this opens
the very intriguing possibility of multiple transitions between these three
types of solutions. \textbf{Option 2} is the second most interesting case
since there is a competition between two types of solutions: the standard
meron $X_{0}=1/2$ and $W_{0}^{2}=v_{0}^{2}$ (type \textbf{I}), and the only
viable type \textbf{II} solution corresponding to the unique positive root of
Eq. (\ref{quadratic}). On the other hand, when \textbf{option 3} is
realized, no transition is possible since the only viable non-Abelian
gravitating solitons belong to type \textbf{I}. In the discussion below, it
will be convenient to introduce the following boundary values of $\Lambda $: 
\begin{equation*}
\Lambda _{1}=\lambda v_{0}^{2}(2e^{2}-\kappa v_{0}^{2})\ ,\qquad \Lambda
_{2}=\frac{2v_{0}^{2}(2e^{2}-\kappa v_{0}^{2})}{e^{2}}\ ,\qquad \Lambda
_{3}=\Lambda _{1}+\frac{e^{2}(2-\lambda e^{2})}{\kappa }\ ,
\end{equation*}

\subsubsection{\textbf{Option 1:} $W_{0}^{2}$ has two positive roots within $
\big(0,v_{0}^{2}\big)$}

The Eq. \eqref{quadratic} has two different positive roots 
\begin{align}
\big( W_{0}^{(\pm)} \big)^{2} = \frac{e^{2}}{\kappa} \big( 1\pm \sqrt{D_{1}} %
\big) \ ,  \label{tworoots}
\end{align}
where 
\begin{align}
D_{1} = 1 + \frac{\kappa \big[ \lambda v_{0}^{2} (\kappa v_{0}^{2} - 2 e^{2}) +
\Lambda \big] }{e^{2} (\lambda e^{2} -2)} \ .
\end{align}
when one of the following sets of conditions is satisfied:

\begin{enumerate}
\item When $0<\lambda<2/e^{2}$: 
\begin{align}
& e^{2}/v_{0}^{2} < \kappa < 2 e^{2}/v_{0}^{2}  \quad \text{and} \quad 
\Lambda_{2} < \Lambda < \Lambda_{3} \ , \\
& \kappa \geq 2e^{2}/v_{0}^{2} \quad \text{and} \quad 
\Lambda_{1} < \Lambda < \Lambda_{3}  \ .
\end{align}

\item When $2/e^{2}<\lambda$: 
\begin{align}
& e^{2}/v_{0}^{2} < \kappa < 2 e^{2}/v_{0}^{2} \quad \text{and} \quad %
 \Lambda_{3} < \Lambda < \Lambda_{2} \ , \\
& \kappa \geq 2e^{2}/v_{0}^{2} \quad \text{and} \quad 
 \Lambda_{3} < \Lambda < \Lambda_{1} \quad \ .
\end{align}
\end{enumerate}

\subsubsection{Option 2: $W_{0}^{2}$ has one positive root in $\big(%
0,v_{0}^{2}\big)$}

The equation \eqref{quadratic} has one positive root $W_{0}^{2} \in \big(0,
v_{0}^{2} \big)$ and one negative root $W_{0}^{2} \in \big(- \infty, 0 \big)$
when the conditions given below are satisfied:

\begin{enumerate}
\item When $0<\lambda<\Lambda/v_{0}^{2} (2e^{2} - \kappa v_{0}^{2})$: 
\begin{align}
& 0 < \kappa < 2 e^{2}/v_{0}^{2} \quad \text{and} \quad 0< \Lambda <
\Lambda_{2} \ , \\
& \kappa > 2e^{2}/v_{0}^{2} \quad \text{and} \quad \Lambda_{2} < \Lambda < 0
\ .
\end{align}

\item When $\lambda > \Lambda/v_{0}^{2} (2e^{2} - \kappa v_{0}^{2})$: 
\begin{align}
& 0 < \kappa < 2 e^{2}/v_{0}^{2} \quad \text{and} \quad \Lambda_{2} <
\Lambda \ , \\
& \kappa > 2e^{2}/v_{0}^{2} \quad \text{and} \quad \Lambda <\Lambda_{2} \ .
\end{align}
\end{enumerate}

In this case, the physical solution is $W_{0} = W_{0}^{(+)}$ given in Eq. %
\eqref{tworoots}.

\paragraph{When $W_{0}^{2}$ has a positive double root in $\big(0,v_{0}^{2} 
\big)$}

The Eq. \eqref{quadratic} has a positive double root 
\begin{equation*}
\big(W_{0}^{(D)}\big)^{2}=\frac{\kappa \lambda
v_{0}^{2}(1+2e^{2})-(4-\lambda )e^{2}}{2\kappa \big[\lambda (1+e^{2})-2\big]}%
\ ,
\end{equation*}%
when the following condition is satisfied: 
\begin{align}
\lambda \neq \frac{2}{e^{2}} \quad \text{and} \quad \kappa > \frac{e^{2}}{%
v_{0}^{2}}
\end{align}%
In this case, the cosmological constant $\Lambda $ and the other coupling
constants are related by 
\begin{equation*}
\lambda (\kappa v_{0}^{2})^{2} - 2 \lambda e^{2} (\kappa v_{0}^{2}) + e^{2}
(2- \lambda e^{2}) + \kappa \Lambda = 0 \ .
\end{equation*}

\subsubsection{\textbf{Option 3:} $W_{0}^{2}$ has no positive roots within $
\big(0,v_{0}^{2}\big)$}

The Eq. \eqref{quadratic} has no positive root within $\big(0,v_{0}^{2}\big)$
when one of the following sets of the conditions is satisfied:

\begin{enumerate}
\item When $0<\lambda< 2/e^{2}$: 
\begin{align}
& 0 < \kappa \leq 2 e^{2}/v_{0}^{2} \quad \text{and} \quad \Lambda \leq
\Lambda_{1} \ , \\
& \kappa > 2e^{2}/v_{0}^{2} \quad \text{and} \quad \Lambda \leq \Lambda_{2}
\ .
\end{align}

\item When $\lambda > 2/e^{2}$: 
\begin{align}
& 0 < \kappa \leq 2 e^{2}/v_{0}^{2} \quad \text{and} \quad \Lambda \geq
\Lambda_{1} \ , \\
& \kappa > 2e^{2}/v_{0}^{2} \quad \text{and} \quad \Lambda \geq \Lambda_{2}
\ .
\end{align}
\end{enumerate}

\begin{figure}[tbp]
\centering
\begin{subfigure}[b]{0.3\textwidth}
        \includegraphics[width=\textwidth]{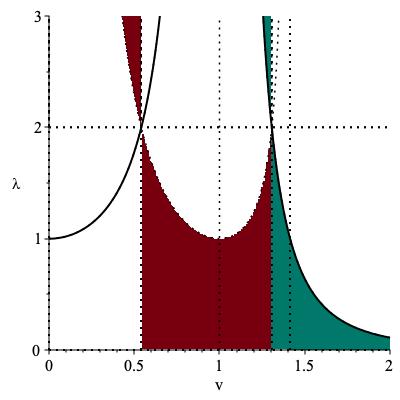}
        \caption{$e=1$, $\Lambda =1$, $\kappa=1$}
        \label{fig:twosoln}
    \end{subfigure}
~ 
\begin{subfigure}[b]{0.3\textwidth}
        \includegraphics[width=\textwidth]{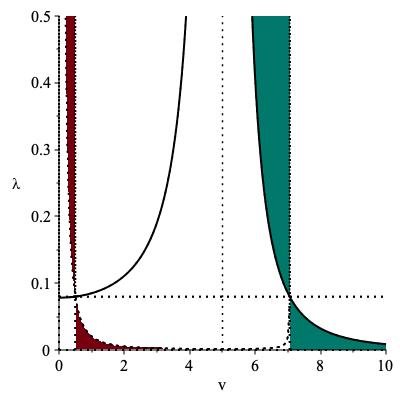}
        \caption{$e=5$, $\Lambda =1$, $\kappa=1$}
        \label{fig:onesoln}
    \end{subfigure}
~ 
\begin{subfigure}[b]{0.3\textwidth}
        \includegraphics[width=\textwidth]{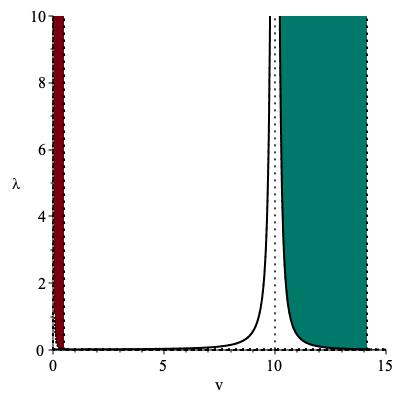}
        \caption{$e=10$, $\Lambda =1$, $\kappa=1$}
        \label{fig:couplings}
    \end{subfigure}
\caption{The burgundy, green, and white regions show the range of couplings
on which Eq. \eqref{quadratic} has one, two, and no positive root of $%
W_{0}^{2}$ in $(0,v_{0}^{2})$, respectively. The solid lines represent
the configurations with double roots. }
\label{fig:solns}
\end{figure}

\subsection{Equation for $y(u,v)$ and Ricci Scalar (both type I and II)}

In all of the above cases, the only non-trivial differential equation is the 
$\theta \theta $-component of the Einstein's equations, which can be written
as 
\begin{equation}
\frac{1}{y^{2}}\partial _{u}\partial _{v}y-\frac{1}{y^{3}}\partial
_{u}y\cdot \partial _{v}y+\frac{1}{y_{0}}=0\ ,
\end{equation}%
where 
\begin{equation}
y_{0}=\Big[\frac{2\kappa }{L^{4}e^{2}F_{0}}X_{0}^{2}(1-X_{0})^{2}-\kappa
\lambda F_{0}(W_{0}^{2}-v_{0}^{2})^{2}-\Lambda F_{0}\Big]^{-1}\ .  \label{y0}
\end{equation}%
The solution to this equation is found to be 
\begin{equation}
y(u,v)=-2C_{2}C_{3}y_{0}\sech^{2}(C_{1}+C_{2}u+C_{3}v)\ .  \label{ysoln}
\end{equation}%
Correspondingly, the Ricci scalar $R_{s}$ is given by 
\begin{equation}
R_{s}=\frac{2}{F_{0}}\Big(\frac{2}{y_{0}}+\frac{1}{L^{2}}\Big)\ .
\label{RicciScalar}
\end{equation}%
The gauge field for both type \textbf{I} and type \textbf{II} are regular
and free of singularity, since the following scalars invariant under
coordinate transformation are found to be 
\begin{equation*}
\text{Tr}\big(F_{\mu \nu }F^{\mu \nu }\big)=-\frac{16X_{0}^{2}(1-X_{0})^{2}}{%
F_{0}^{2}L^{4}}\ ,\qquad \text{Tr}\big(A_{\mu }A^{\mu }\big)=-\frac{%
4X_{0}^{2}}{F_{0}L^{2}}\ .
\end{equation*}%
One can also find that the space-time of this solution belongs to the Petrov
type D.

\section{Thermodynamics in Nariai Coordinates}

Without loss of generality, one can rescale the coordinates $u$ and $v$ to
choose the constants $C_{1}=0$, $C_{2}=1$, and $C_{3}=-1$. Then, the metric
becomes 
\begin{equation}
ds^{2}=-2F_{0}y_{0}\sech^{2}(u-v)dudv+F_{0}L^{2}d\Omega _{2}^{2}\ .
\label{metric2}
\end{equation}%
Introducing new coordinates $\tilde{t}$ and $R$ given by 
\begin{equation*}
u=\frac{\tilde{t}+R}{2}\ ,\qquad v=\frac{\tilde{t}-R}{2}\ ,
\end{equation*}%
we can express the metric \eqref{metric2} becomes 
\begin{equation}
ds^{2}=\frac{F_{0}y_{0}}{2}\sech^{2}(R)\big(-d\tilde{t}^{2}+dR^{2}\big)%
+F_{0}L^{2}d\Omega _{2}^{2}\ .  \label{metric3}
\end{equation}%
Again, introducing new coordinates $t$ and $r$ given by 
\begin{equation*}
t=r_{0}\tilde{t}\ ,\qquad \sech^{2}(R)\equiv 1-\frac{r^{2}}{r_{0}^{2}}\ 
\end{equation*}%
the metric \eqref{metric3} can be transformed to the following metric
representing a product space $dS_{2}\times S^{2}$: 
\begin{equation}
ds^{2}=\frac{F_{0}y_{0}}{2r_{0}^{2}}\Big[-\Big(1-\frac{r^{2}}{r_{0}^{2}}\Big)%
dt^{2}+\frac{1}{1-\frac{r^{2}}{r_{0}^{2}}}dr^{2}\Big]+F_{0}L^{2}d\Omega
_{2}^{2}\ .  \label{NariaiCoord}
\end{equation}%
This metric clearly shows that our space-time belongs to the Nariai family
of space-times. With this coordinate system, it is straightforward to
compute the surface gravity $\kappa $ 
\begin{equation*}
\kappa =\frac{1}{r_{0}}\ ,
\end{equation*}%
so that the temperature $T$ is found to be 
\begin{equation*}
T=\frac{1}{2\pi r_{0}}\ .
\end{equation*}%
It should be noticed that (as it will be shown in the next section) the
entropies of the two different families are inversely proportional to $%
r_{0}^{2}$, so that the ratio of the entropies (which determines which
family prevails) is independent of $r_{0}^{2}$. In the next section, 
we will compute the entropy functions of type \textbf{I} and type 
\textbf{II} and find the range of the physical parameters where 
one of these type dominates the other.

\section{Thermodynamics and Transitions between families}

One of the most interesting outcomes of the present analysis is that there
are different types of gravitating non-Abelian solitons in the
Einstein-Yang-Mills-Higgs in the sector described above. The space-time
geometry is of Nariai type, while the Yang-Mills field can be either a
standard ($X_{0}=1/2$) meron with a trivial Higgs field or a non-standard
meron with a non-trivial Higgs field. Hence, the natural question is: 
\textit{which one of the gravitating solitons described above will prevail?}
The analysis shows that the answer to this question depends in a crucial way
both on the Higgs coupling and on the ``VEV" of the Higgs field itself. There
are two related tools to answer this question: the first is the computation
of the ``entropy" of the solutions, the second is the computation of the
Euclidean action of the different configurations (as it will be shown, these
tools give consistent answers).

\subsection{Entropy function}

The entropy function(al) of space-times of the form $AdS_{2}\times S^{D-2}$
(associated to the near-horizon geometries of extremal black holes) for an
arbitrary dimension $D$ was studied by A. Sen \cite{Sen2005}. The entropy
function of this class of space-times is the product of $2\pi $ with the
Legendre transform of the Lagrangian density integrated over $S^{D-2}$.
Since the electric fields associated with the gauge fields play the role of
configuration variables, this entropy is a Routhian density over $AdS_{2}$
rather than a Hamiltonian density. A direct application of the Sen's method
to Nariai class shows that the entropy of a Nariai space-time is
\textquotedblleft minus" Routhian density over $dS_{2}$ \cite{Cho2008}. The
first step corresponds to the on-shell evaluation of the Lagrangian of the
matter field (while the second step corresponds to integrate the on-shell
Lagrangian for the matter field over $S^{2}$). The on-shell Lagrangian reads 
\begin{align}
\frac{1}{4e^{2}}\text{Tr}\big[F_{\mu \nu }F^{\mu \nu }\big]& =-\frac{1}{%
4e^{2}}\Big(\frac{4X_{0}(1-X_{0})}{F_{0}L^{2}}\Big)^{2}\ ,  \notag \\
\frac{1}{2e^{2}}\text{Tr}\big[D_{\mu }\Phi ^{\ast }D^{\mu }\Phi \big]& =-%
\frac{1}{2e^{2}}\Big(\frac{2W_{0}(2X_{0}-1)}{F_{0}L^{2}}\Big)^{2}\ ,
\end{align}%
and the Ricci scalar was given by \eqref{RicciScalar}. In the Nariai
coordinates, we have 
\begin{equation*}
\sqrt{-g}=\frac{F_{0}^{2}y_{0}L^{2}}{2r_{0}^{2}}\sin \theta \ .
\end{equation*}%
Since the system has no electric field, the Routhian density $H$ becomes 
\begin{eqnarray}
H &=&-\oint_{S^{2}}d\theta d\phi \sqrt{-g}\,\mathcal{L}  \notag \\
&=&-\frac{2\pi F_{0}^{2}y_{0}L^{2}}{r_{0}^{2}}\Big[\frac{2}{F_{0}}\Big(\frac{%
2}{y_{0}}+\frac{1}{L^{2}}\Big)-\frac{1}{4e^{2}}\Big(\frac{4X_{0}(1-X_{0})}{%
F_{0}L^{2}}\Big)^{2}-\frac{1}{2e^{2}}\Big(\frac{2W_{0}(2X_{0}-1)}{F_{0}L^{2}}%
\Big)^{2}-\lambda (W_{0}^{2}-v_{0}^{2})^{2}\Big]\ ,
\end{eqnarray}%
where $\mathcal{L}$ is the Lagrangian of the system. The computations given
in \cite{Sen2005, Cho2008} tells us that the entropy function S will be
given by 
\begin{equation}
S=-2\pi H_{\text{max}}\ .  \label{sdef}
\end{equation}%
where $H_{\text{max}}$ is the maximum value with respect to $F_{0}y_{0}$ and 
$F_{0}L^{2}$ (note that $H_{\text{max}}$ is negative so that the entropy is
positive definite as it should). The entropy function of the system is found
to be 
\begin{equation*}
S=%
\begin{cases}
\dfrac{4\pi ^{2}}{\kappa e^{2}r_{0}^{2}\big(\kappa \lambda
(W_{0}^{2}-v_{0}^{2})^{2}+\Lambda \big)}\Big(e^{2}-\kappa
W_{0}^{2}(2X_{0}-1)^{2}+\sqrt{Q}\Big) & \text{for}\ W_{0}^{2}\neq v_{0}^{2}%
\vspace{0.2in} \\ 
\dfrac{4\pi ^{2}\big(1+\sqrt{1-\kappa \Lambda /2e^{2}}\big)}{\kappa \Lambda
r_{0}^{2}} & \text{for}\ W_{0}^{2}=v_{0}^{2}\ .%
\end{cases}%
\end{equation*}%
where 
\begin{equation*}
Q=\big(e^{2}-\kappa W_{0}^{2}(2X_{0}-1)^{2}\big)^{2}-8\kappa
e^{2}X_{0}^{2}(1-X_{0})^{2}\big(\kappa \lambda
(W_{0}^{2}-v_{0}^{2})^{2}+\Lambda \big)\ .
\end{equation*}

\subsection{Euclidean action}

In order to compute the Euclidean action for both families, it is convenient
to define the "sizes" of the $dS_{2}$ and $S^{2}$ as 
\begin{equation*}
v_{1}=F_{0}y_{0}\ ,\qquad v_{2}=F_{0}L^{2}\ ,
\end{equation*}%
respectively. Then, the Euclidean action can be written as 
\begin{equation*}
I_{E}=\frac{4\pi ^{2}v_{1}v_{2}}{r_{0}^{2}}\Big[2\Big(\frac{1}{v_{1}}+\frac{1%
}{v_{2}}\Big)-\frac{1}{4e^{2}}\Big(\frac{4X_{0}(1-X_{0})}{v_{2}}\Big)^{2}-%
\frac{1}{2e^{2}}\Big(\frac{2W_{0}(2X_{0}-1)}{v_{2}}\Big)^{2}-\lambda
(W_{0}^{2}-v_{0}^{2})^{2}\Big]\ .
\end{equation*}%
The on-shell conditions that extremize the Euclidean action are found to be 
\begin{align}
2\lambda e^{2}(W_{0}^{2}-v_{0}^{2})v_{2}+(2X_{0}-1)^{2}& =0\ , \\
(2X_{0}-1)\big[W_{0}^{2}v_{2}-X_{0}(1-X_{0})\big]& =0\ , \\
e^{2}\big[\kappa \lambda (v_{0}^{2}-W_{0}^{2})^{2}+\Lambda \big]v_{2}^{2}+%
\big[\kappa (2X_{0}-1)^{2}W_{0}^{2}-e^{2}\big]v_{2}+2\kappa
X_{0}^{2}(1-X_{0})^{2}& =0\ , \\
\big[e^{2}\big(\kappa \lambda (v_{0}^{2}-W_{0}^{2})^{2}+\Lambda \big)%
v_{2}^{2}-2\kappa X_{0}^{2}(1-X_{0})^{2}\big]v_{1}+e^{2}v_{2}& =0\ 
\end{align}%
One can easily check that these equations are equivalent to the equations of
motion \eqref{algebr1}, \eqref{algebr2}, \eqref{reduced1}, and \eqref{y0}.
Thus, it follows from \eqref{sdef} that the on-shell Euclidean action is
precisely equal to the entropy of the system. The overall factor $2\pi $
should be understood as the circumference of the imaginary time.

\subsection{Useful Plots}

The previous analysis showed that a critical parameter to determine which of
the solutions is thermodynamically favored is the VEV of the Higgs field $%
v_{0}^{2}$. Here below, we will include the plots which clarify the
comparisons between the two families of gravitating solitons in the three
options defined in the previous sections (depending on the number of roots
in the equation for $W_{0}^{2}$).

\subsubsection{Option 1 Plots}

In the ``\textit{Option 1 case}" defined in the previous sections, Eq. %
\eqref{quadratic} has two different positive real roots (let us call them $%
\left( W_{0}^{2}\right) ^{\pm }$ where the $+$ stands for the larger root and
the $-$ for the smaller one). In this case, multiple transitions may appear as
the thermodynamics is determined by the comparison of three solutions: the
type \textbf{I}, the type \textbf{II} with root $\left( W_{0}^{2}\right)
^{+} $ and the type \textbf{II} with root $\left( W_{0}^{2}\right) ^{-}$.
These three solutions will be characterized by their own entropy (Euclidean
actions): let us call $S_{I}$, $S_{II}^{+}$ and $S_{II}^{-}$ the entropy of
the type \textbf{I} solution, of the type \textbf{II} solution with root $%
\left( W_{0}^{2}\right) ^{+}$ and the type \textbf{II} solution with root $%
\left( W_{0}^{2}\right) ^{-}$ respectively. Obviously, $S_{I}$, $S_{II}^{+}$
and $S_{II}^{-}$ \ (which have been constructed explicitly in the previous
subsection) depends on all the parameters of the model $\lambda $, $e$, $%
\Lambda $, and so on. Here we will emphasize especially the dependence on
the VEV of the Higgs field $v_{0}^{2}$ as $v_{0}^{2}$ appears to be quite
crucial to determine the phases of the system.

Hence, using the results from the previous subsection, we get%
\begin{equation}
S_{I}=S_{I}\left( x\right) = \dfrac{4\pi ^{2} \big( 1 + \sqrt{1- \kappa
\Lambda / 2 e^{2}} \big) } {\kappa \Lambda r_{0}^{2}} \ ,  \label{typeIplot}
\end{equation}%
\begin{equation}
S_{II}^{+}=S_{II}^{+}\left( x\right) = \frac{4 \pi^{2} \big[ e \big( 2 +
\lambda e^{2 }D_{1} -( \lambda \kappa x- \lambda e^{2} -2 ) \sqrt{D_{1}}\big)%
+\sqrt{f_{+}(x)} \big] }{e r_{0}^{2} \big[ \lambda \kappa x + (2-\lambda
e^{2})(1 + \sqrt{D_{1}}) \big] \big[ \lambda ( e^{2} (1+ \sqrt{D_{1}}) -
\kappa x )^{2}+ \kappa \Lambda \big] } \ ,  \label{typeIIplota}
\end{equation}%
\begin{equation}
S_{II}^{-}=S_{II}^{-}\left( x\right) = \frac{4 \pi^{2} \big[ e \big( 2 +
\lambda e^{2 }D_{1} +( \lambda \kappa x- \lambda e^{2} -2 ) \sqrt{D_{1}}\big)%
+\sqrt{f_{-}(x)} \big] }{e r_{0}^{2} \big[ \lambda \kappa x + (2-\lambda
e^{2})(1 - \sqrt{D_{1}}) \big] \big[ \lambda ( e^{2} (1- \sqrt{D_{1}}) -
\kappa x )^{2}+ \kappa \Lambda \big] } \ ,  \label{typeIIplotb}
\end{equation}%
where 
\begin{align}
& f_{\pm}(x) = e^{2} (\lambda e^{2} -2) (\lambda e^{2} D_{1} -2) (1 \pm 
\sqrt{D_{1}})^{2} + \kappa \lambda x \big[ \kappa ( \lambda e^{2} D_{1} - 2 %
\big(1 \pm \sqrt{D_{1}})^{2} \big) x  \notag \\
& \hspace{.5in} - 2 e^{2} \big(1 \pm \sqrt{D_{1}}) \big( (\lambda e^{2} -2)
D_{1} - 2(1 \pm \sqrt{D_{1}}) \big) \big] - 2 \kappa \Lambda \big(1 \pm 
\sqrt{D_{1}})^{2} \ .
\end{align}

\begin{equation*}
x=v_{0}^{2}\ .
\end{equation*}%
Here we plot together $S_{I}$, $S_{II}^{+}$ and $S_{II}^{-}$ as
function of $x$ in three different cases (\textit{case 1}: $e=100$, 
$\lambda=1$; \textit{case 2}: $e=1$, $\lambda =1$; \textit{case 3}: 
$e=1$, $\lambda =100$) in order to show the
differences between the situations in which the Higgs and Yang-Mills
coupling are equal, and one small and one large. In these
three plots, both $\Lambda $\ and $\kappa $\ will be fixed to 1. As one can
see from these figures, the most preferred configuration is determined 
in a sensitive way depending on the physical parameters. For the 
parameters used in these three figures, the most favorable 
configuration is the standard meron. The last figure shows that 
$S_{II}^{-}$ wins $S_{II}^{+}$ for some certain values of the 
parameters. 
\begin{figure}[tbp]
\centering
\begin{subfigure}[b]{0.3\textwidth}
        \includegraphics[width=\textwidth]{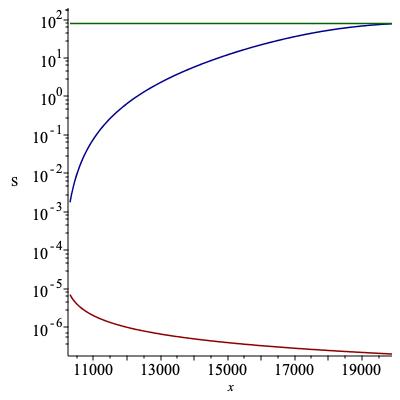}
        \caption{$e=100$, $\lambda=1$}
        \label{fig:tworoots1}
    \end{subfigure}
\begin{subfigure}[b]{0.3\textwidth}
        \includegraphics[width=\textwidth]{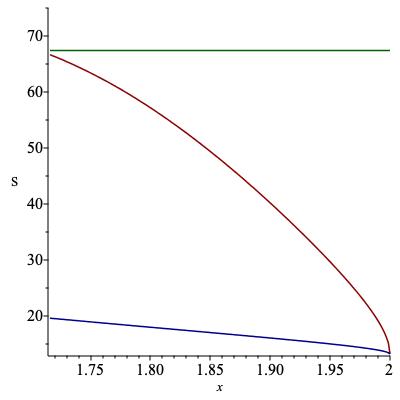}
        \caption{$e=1$, $\lambda=1$}
        \label{fig:tworoots2}
    \end{subfigure}
~ 
\begin{subfigure}[b]{0.3\textwidth}
        \includegraphics[width=\textwidth]{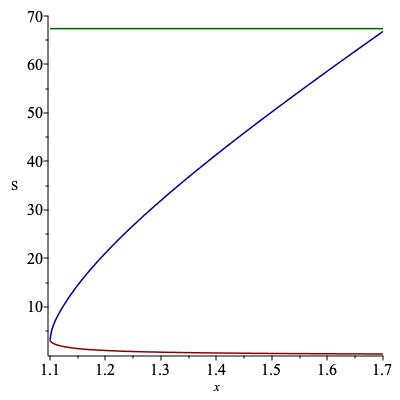}
        \caption{$e=1$, $\lambda =100$}
        \label{fig:tworoots3}
    \end{subfigure}
\caption{The entropy functions with $\protect\kappa =\Lambda =r_{0}=1$. The
green, red, and blue curves are the graphs of $S_{I}$, $S_{II}^{+}$, and $%
S_{II}^{-}$, respectively.}
\label{fig:tworoots}
\end{figure}

\subsubsection{Option 2 Plots}

In the "\textit{Option 2 case}" defined in the previous sections, Eq. %
\eqref{quadratic} has one positive real roots (let us call it just $%
W_{0}^{2} $). In this case, transitions may appear as the thermodynamics is
determined by the comparison of two solutions: the type \textbf{I} and the
type \textbf{II} with root $W_{0}^{2}$. These two solutions will be
characterized by their own entropy (Euclidean actions): let us call $S_{I}$
and $S_{II}$ the entropy of the type \textbf{I} solution and of the type 
\textbf{II} solution respectively. As in the option 1 case, $S_{I}$ and $%
S_{II}$ depends on all the parameters of the model, but we will emphasize the
dependence on $v_{0}^{2}$.

\begin{figure}[tbp]
\centering
\begin{subfigure}[b]{0.3\textwidth}
        \includegraphics[width=\textwidth]{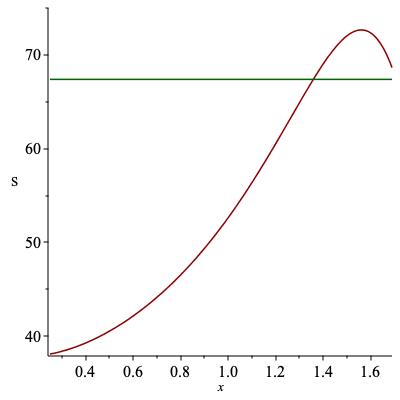}
        \caption{$e=1$, $\lambda=1$}
        \label{fig:oneroot1}
    \end{subfigure}
~ 
\begin{subfigure}[b]{0.3\textwidth}
        \includegraphics[width=\textwidth]{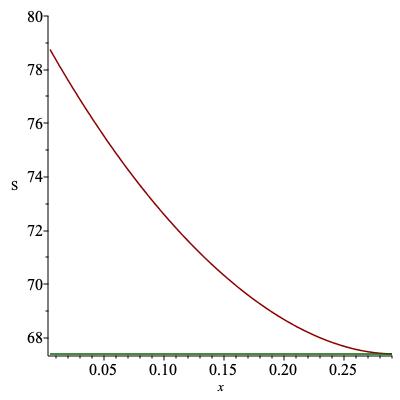}
        \caption{$e=1$, $\lambda =100$}
        \label{fig:oneroot2}
    \end{subfigure}
~ 
\begin{subfigure}[b]{0.3\textwidth}
        \includegraphics[width=\textwidth]{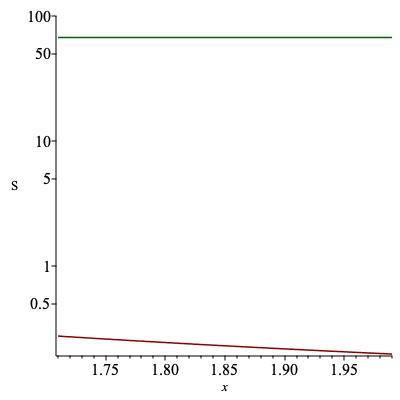}
        \caption{$e=1$, $\lambda =100$}
        \label{fig:oneroot3}
    \end{subfigure}
\caption{The entropy functions with $\protect\kappa =\Lambda =r_{0}=1$. The
green and red curves are the graphs of $S_{I}$ and $S_{II}$, respectively. The 
last two figures show two different regions for the same physical parameters. 
For the case with $e=1$ and $\lambda=100$, the interval $0.29<x<1.71$ is 
forbidden. In Fig. \ref{fig:oneroot1}, the type \textbf{I} is favored for 
$x<1.359$ and the type \textbf{II} is favored for $x>1.359$. In Fig. \ref{fig:oneroot2}, 
the type \textbf{II} always prevails, and in Fig. \ref{fig:oneroot3}, the type \textbf{I}
is always preferred. }
\label{fig:oneroot}
\end{figure}

Using the results from the previous subsection, we get the entropy function
of the type \textbf{II} with a double root 
\begin{equation}
S_{II}^{\text{(D)}}=S_{II}^{\text{(D)}}\left( x\right) =\frac{4\pi ^{2}}{%
er_{0}^{2}\big[\lambda \kappa x+2-\lambda e^{2})\big]}\ .
\label{option2plotII}
\end{equation}%
The entropy function for the case with one positive and one negative 
roots equals to $S_{II}^{+}$.
Also in this case, we plot together $S_{I}$ and $S_{II}$ as function of 
$x$ in two different cases (\textit{case 1}: $e=1$, $\lambda =1$; 
\textit{case 2}: $e=1$, $\lambda =100$) in order to show the
differences between the situations in which the Higgs and Yang-Mills
coupling are equal, and $\lambda$ is larger than $e$. In these
three plots, $r_{0}$, $\Lambda $, and $\kappa $ will be fixed to 1. 
It should be emphasized that for a given set of the parameters 
$\{ e, \lambda, \kappa, \Lambda, r_{0} \}$, the most preferred 
configuration can change as the square of the VEV varies. Interestingly,
such a configuration changes continuously in Fig. (\ref{fig:oneroot1}) 
whereas it changes in a discontinuous way, as can be seen in Fig.
(\ref{fig:oneroot2}) and (\ref{fig:oneroot3}).

\section{Spin from Isospin and 4DQHE}

The analysis of the Klein-Gordon equation reveals some crucial differences
between the present non-Abelian versions of Nariai space-times and the well
known Nariai space-times, which are solutions of the Einstein-Maxwell field
equations \cite{Bertotti, Robinson}.

Let us begin this section with a very short review of the \textit{%
spin-from-isospin effect} \cite{spinisospin1} \cite{spinisospin2} \cite%
{spinisospin3}. Roughly speaking, in the case of topologically non-trivial
configurations that are not spherically symmetric but whose energy-momentum
tensor is spherically symmetric the lack of spherical symmetry under spatial
rotation is compensated by an internal transformation. This leads to a
modification of the definition of the angular momentum, which in the usual
example of non-Abelian $SU(2)$ monopoles reads:%
\begin{equation}
\overrightarrow{L}\rightarrow \overrightarrow{J}=\overrightarrow{L}-\frac{1}{%
2}\overrightarrow{\tau }\ ,  \label{modJ}
\end{equation}%
where $\overrightarrow{L}$\ is the orbital angular momentum and $%
\overrightarrow{\tau }$ are the $SU(2)$ generators entering in the ansatz of
the gauge field. The new term in the definition of $\overrightarrow{J}$ is
exactly related to the infinitesimal internal rotation needed to compensate
the lack of spherical symmetry under spatial rotation. As it has been
discussed in the original references \cite{spinisospin1} \cite{spinisospin2} 
\cite{spinisospin3} this leads scalar test fields moving in the background
of this type of gauge fields to behave as Fermions. From the viewpoint of
the Klein-Gordon equation this can be directly seen as follows. The
non-Abelian flat Klein-Gordon operator reads (see \cite{mimeron1} \cite%
{mimeron3})%
\begin{equation}
ds^{2}=-dt^{2}+dr^{2}+r^{2}\left( d\theta ^{2}+\sin ^{2}\theta d\phi
^{2}\right) \ ,  \label{usualmetric}
\end{equation}%
\begin{equation*}
D_{\mu }D^{\mu }=\left( \nabla _{\mu }-\left[ A_{\mu },\cdot \right] \right)
\left( \nabla ^{\mu }-\left[ A^{\mu },\cdot \right] \right) \Rightarrow
\end{equation*}%
\begin{equation}
D_{\mu }D^{\mu }=\square _{2D}-\frac{\left( \overrightarrow{J}\cdot 
\overrightarrow{J}-c_{0}\right) }{r^{2}},  \label{spinfromisospinsimple}
\end{equation}%
where $\square _{2D}$ is the two-dimensional D'Alembertian in the $t$ and $r$
coordinates and $c_{0}$ is a constant which depends on the representation of
the test scalar field. Hence, Eq. (\ref{spinfromisospinsimple}) explains in
a very simple way the effect of the ``need to compensate" the lack of
spherical symmetry with an internal rotation: the centrifugal barrier (which
is the term that goes as $1/r^{2}$ for large $r$) is modified. Obviously,
it is precisely from the centrifugal-barrier like term that one usually
reads the spin of the ``test fields". A quite trivial (but useful as we will
now show) observation is the following: the $1/r^{2}$ factor which
multiplies $\overrightarrow{J}\cdot \overrightarrow{J}$ arises because of
the $r^{2}$ term in front of the two-sphere in the metric in Eq. (\ref%
{usualmetric}). Now, we are ready for an important question:

\textit{What happens if in the spherically symmetric space-time of our
interest (sourced by a non-Abelian soliton) in front of the two-sphere we
have just a constant instead of the coordinate-dependent factor} $r^{2}$?
What happens to the \textit{spin-from-isospin} effect?

Eq. (\ref{spinfromisospinsimple}) suggests an intriguing answer: the
(modified) centrifugal barrier becomes a term that does not depend on $r$
at all: such a term possesses discrete degenerate energy levels (as it is
proportional to $\overrightarrow{J}\cdot \overrightarrow{J}$) with an energy
gap proportional to the (homogeneous) magnetic flux. In other words, when 
\textit{in front of the two-sphere we have a constant instead of} $r^{2}$
the ``spin-from-isospin" term is replaced by the typical Hamiltonian which is
used to describe the QHE in higher dimensions\footnote{%
Although this is (to a certain extent) not too surprising taking into
account that the present gravitating solitons possess a non-trivial
(non-Abelian) magnetic flux, the fact that there are two families of
solitons competing against each other lead to very interesting consequences.%
} (see in particular, \cite{4DQHE1} \cite{4DQHE2}). We will now show that
this is indeed the case, and that the degeneracy of the energy levels of the
non-Abelian Klein-Gordon equation changes dramatically when passing from the
type \textbf{I} to the type \textbf{II} solutions: such an effect is a
genuine non-Abelian fingerprint of the present families of gravitating
solitons.

Let us consider scalar test fields $\Psi ^{a}$ $(a=1,2)$, which transforms
like a two-component vector, in our background space-time. The field
equation is given by 
\begin{equation*}
D^{2}\Psi -m^{2}=0\ ,
\end{equation*}%
where $m$ is the mass of the scalar fields. Note that the Yang-Mills field
associated with our solution satisfies 
\begin{equation*}
\nabla ^{\mu }A_{\mu }=0\ ,\qquad A^{\mu }A_{\mu }=-\frac{2X_{0}^{2}}{%
L^{2}F_{0}}\mathbf{1}_{2\times 2}\ .
\end{equation*}%
The Klein-Gordon equation can be written as 
\begin{equation*}
\nabla ^{2}\Psi -m^{2}\Psi +2A^{\mu }\nabla _{\mu }\Psi +A^{\mu }A_{\mu
}\Psi =0\ .
\end{equation*}%
Explicitly, it can be written as 
\begin{equation*}
\Delta _{(u,v)}\Psi +\Big[\frac{1}{F_{0}L^{2}}\big(\vec{L}-X_{0}\vec{\tau}%
\big)^{2}+\frac{X_{0}^{2}}{F_{0}L^{2}}-m^{2}\Big]\Psi =0\ ,
\end{equation*}%
where $\hat{L}=\hat{L}_{i}\tau _{i}$ for the standard angular momentum
operators $\hat{L}_{i}$, and $\Delta _{(u,v)}$ is the 2 dimensional
D'Alembertian operator given by 
\begin{equation*}
\Delta _{(u,v)}=-\frac{2}{F_{0}y_{0}}\partial _{u}\partial _{v}\ .
\end{equation*}%
The eingenfunction of this operator has the form of
\begin{align}
\Psi = e^{au+bv} \Psi_{0} \ ,
\end{align}
with the eigenvalue $-2ab/F_{0}y_{0}$, for arbitrary constants $a$ and $b$,  
and a constant doublet $\Psi_{0}$.

\subsection{Differences between type I and type II non-Abelian solitons}

In the case of the type I non-Abelian gravitating solitons one can compute
the sum of the orbital and isospin angular momenta through a standard
procedure. In particular, the eigenvalues of the spin-orbit coupling can be
obtained by 
\begin{equation}
\vec{J}^{2}=\Big(\vec{L}+\frac{1}{2}\vec{\tau}\Big)^{2}=\vec{L}^{2}+\frac{1}{%
4}\vec{\tau}^{2}+\vec{L}\cdot \vec{\tau}\quad \Longrightarrow \quad \vec{L}%
\cdot \vec{\tau}=\vec{J}^{2}-\vec{L}^{2}-\frac{1}{4}\vec{\tau}^{2}
\label{spinmanipulation}
\end{equation}%
which gives the eigenvalues of the coupling operator $\vec{L}\cdot \vec{\tau}
$ 
\begin{equation*}
j(j+1)-l(l+1)-\frac{1}{2}\cdot \Big(\frac{1}{2}+1\Big)\ .
\end{equation*}%
Thus, the part of the Hamiltonian which results in a spin-isospin effect 
\begin{equation}
H_{\text{spin-isospin}}=\frac{1}{F_{0}L^{2}}\big(\vec{L}-X_{0}\vec{\tau}\big)%
^{2}
 \label{sfromi}
\end{equation}%
has the same degeneracy of the physical system as the usual spin-orbit
couplings when $X_{0}=1/2$. It gives a very similar behavior\footnote{%
Indeed, as it happens in \cite{4DQHE1}, also in the present case in order
to increase the degeneracy of the discrete energy levels of the Hamiltonian
we must increase the dimension of the representation of the test field.} of
the Hamiltonian of the four dimensional quantum hall effect proposed in \cite%
{4DQHE1}, 
\begin{equation*}
H_{\text{QHE}}=\frac{1}{2MR^{2}}\sum_{a<b}\Lambda _{ab}^{2}\ ,
\end{equation*}%
where $\Lambda _{ab}=-i\big(x_{a}D_{b}-x_{b}D_{a}\big)$. Here, $F_{0}L^{2}$,
the denominator of \eqref{sfromi}, plays the same role as the radius
of the orbit in the system with QHE. \newline

On the other hand, in the gravitating solitons of type \textbf{II} (In the
case with $X_{0}\neq 1/2$ and a non-trivial Higgs field) the (\textit{would
be}) total angular momentum $\overrightarrow{J}$ becomes 
\begin{equation*}
\overrightarrow{J}=\vec{L}-X_{0}\vec{\tau}
\end{equation*}%
with a real coefficient $X_{0}$ different from $1/2$ (as, generically, $%
X_{0} $ is not even a rational number: see Eqs. (\ref{TypeIIX0}) and (\ref%
{quadratic})). The most dramatic effect manifests itself in the degeneracy
of the energy level associated to the operator \eqref{sfromi}.
This can be understood easily looking at the standard manipulations in Eq. (%
\ref{spinmanipulation}) in the case in which $X_{0}$ is a generic real
number: 
\begin{eqnarray}
\overrightarrow{J} =\vec{L}-X_{0}\vec{\tau}\ ,\ and\ \ X_{0}\neq 1/2,\ 0
\qquad \Longrightarrow \qquad
\lbrack J_{i},J_{k}] \neq \varepsilon _{ikl}J_{l}.  \label{sfromi2}
\end{eqnarray}%

Although one may still hope to find a rigorous definition of the ``total
angular momentum $\overrightarrow{J}$" with $X_{0}$ a real number, it is
clear that \textit{one should not expect} that the spectrum of the above
``spin-from-isospin operator" $H_{\text{spin-isospin}}$ in Eq. (\ref{sfromi})
is still related to the combination%
\begin{equation*}
j\left( j+1\right) \ ,\ 
\end{equation*}
\begin{equation*}
j=l+X_{0}\ .
\end{equation*}%
The reason is that Eq. (\ref{sfromi2}) suggests that the\ eigenvalues of $H_{%
\text{spin-isospin}}$ depend on $l$ (related to the eigenvalues of $\vec{L}$%
) and $s$ (related to the eigenvalues of $\vec{\tau}$) \textit{separately}
(and therefore the degeneracy is expected to be lower than in the case with $%
X_{0}=1/2$ when the eigenvalues only depend on $j$). Consequently, the
non-Abelian Klein-Gordon equation associated to the type \textbf{II}
gravitating merons should have energy levels with a different degeneracy
than in the case of the type \textbf{I} gravitating solitons.

This has the following potential consequence. A multi-Fermionic system
(charged under the gauge group) living in the type \textbf{I} gravitating
solitons (in the approximation in which these Fermions can be considered as
test fields) would be subject to a Hamiltonian with many of the features of
the 4D QHE (as it has been explained here above). The same multi-Fermionic
system would perceive a Hamiltonian with very different degeneracies in the
type \textbf{II} gravitating solitons. Therefore, if there is a
semiclassical transition from one family to the other\footnote{%
This could happen, for instance, if there is a slight change in the VEV of
the Higgs field around a value at which the family \textbf{II} starts to
overcome the family \textbf{I}.} the multi-Fermionic system would suddenly
be subject to a different QHE-like Hamiltonian with completely different
degeneracies. We hope to come back on the fascinating physical properties of
these scenarios in a future publication.

\section{Conclusions}

In this article, we constructed the first two analytic families of globally
regular non-Abelian gravitating solitons in the Einstein-Yang-Mills-Higgs
theory in (3+1)-dimensions with the Higgs field in the adjoint
representation (however, the case in which the Higgs field is in
the fundamental is very similar). The space-time metric is of Nariai
type in both cases. The Yang-Mills fields are of meron type (namely,
proportional to a pure gauge: $A_{\mu }=X_{0}U^{-1}\partial _{\mu }U$ for
some parameter $X_{0}$) for both families. On the other hand, while in the
first family (called type \textbf{I} in the main text) of non-Abelian
gravitating soliton $X_{0}=1/2$ (as in all the known examples of merons
available so far) and the Higgs field is trivial, in the second family (type 
\textbf{II}) $X_{0}\neq 1/2$ and the Higgs field is non-trivial (to the best
of the authors knowledge, this is the first example of meron solutions with $%
X_{0}\neq 1/2$). We have compared these two families of globally regular
gravitating solitons by computing the Euclidean action of both types. This
allows to determine when type \textbf{II} solitons (with a non-trivial Higgs
and $X_{0}\neq 1/2$) are favored over type \textbf{I} solitons and \textit{%
viceversa}. It turns out that the most favored configuration is determined
in a sensitive way depending on the parameters of the model. Even for 
a given set of the parameters other than $x=v_{0}^{2}$, the most 
preferred configuration changes continuously or discontinuously as 
$x$ varies. In order to disclose the differences between
type \textbf{I} and type \textbf{II} gravitating solitons we analyzed the
non-Abelian Klein-Gordon equation for a test scalar field minimally coupled
to the non-Abelian fields sourcing the gravitating solitons themselves. The
Klein-Gordon equation is able to detect very neatly the difference between
type \textbf{I} and type \textbf{II} solitons (despite the fact that the
space-time metric is similar in both cases). The Klein-Gordon equation can
be written as the sum of a two-dimensional D`Alambert operator plus one of
the Hamiltonians which has been proposed in the literature to describe the
four-dimensional Quantum Hall Effect (QHE): the difference between type 
\textbf{I} and type \textbf{II} solutions manifests itself in a difference
between the degeneracies of the corresponding energy levels. This opens the
very intriguing perspective to analyze the ``many-body" wave functions of
multi-Fermionic systems minimally coupled to the regular meronic fields of
type \textbf{I} and type \textbf{II} solutions (in the test field limit in
which these Fermions do not modify substantially the space-time metric). The
idea would be to distinguish type \textbf{I} and type \textbf{II} solutions
by looking at the degeneracy of the corresponding Landau Levels. We will
return to this very interesting issue in a future publication.

\bigskip

\section*{ Acknowledgment}

F. C. has been funded by Fondecyt Grants 1200022. The Centro de Estudios
Cient\'{\i}ficos (CECs) is funded by the Chilean Government through the
Centers of Excellence Base Financing Program of ANID. S. H. O. is supported
by the National Research Foundation of Korea funded by the Ministry of
Education of Korea (Grant 2018-R1D1A1B0-7048945, 2017- R1A2B4010738).

\end{document}